\documentclass[12pt,a4paper]{article}

\usepackage[english]{babel}
\usepackage{amsmath,amssymb,amsfonts,amsthm}
\usepackage[mathscr]{eucal}
\usepackage[all]{xy}
\usepackage{hyperref}
\usepackage{setspace}
\usepackage{upgreek}
\usepackage{hyperref}
\usepackage{cite}

\usepackage{times}
\usepackage{color}

\usepackage{indentfirst}

\allowdisplaybreaks

\begin{document}

\title{On the world sheet of anyon in the external electromagnetic field}
\author {D.S.~Kaparulin and I.A.~Retuntsev}
\date{\small\textit{
Physics Faculty, Tomsk State University, Lenina ave. 36, Tomsk
634050, Russia}}

\maketitle

\begin{abstract}
\noindent We study the issue of description of spinning particle
dynamics by means of recently proposed world sheet concept. A model
of irreducible spinning particle in the $3d$ Minkowski space with
two gauge symmetries is considered. The classical trajectories of
free particle lie on a circular cylinder with a time-like axis. The
direction of cylinder axis is determined by the particle momentum,
and its position in space-time depends on the value of total angular
momentum. The radius of cylinder is determined by the
representation. The model admits inclusion of consistent
interactions with a general (not necessarily uniform)
electromagnetic field. The classical trajectories of the particle
lie on the cylindrical hypersurface in space time, whose position is
determined by the initial values of the momentum and total angular
momentum. All the world paths that lie on one and the same
representative in the set of hypersufaces are connected by gauge
transformations. The general construction is illustrated by the
example of unform electric field. In this case, the particle paths
are shown to be general pseudotoroidal lines.
\end{abstract}

\section{Introduction}

The dynamics of particles with an internal angular momentum is
studied for century. Frenkel has been the first to describe motion
of charged particle with spin in an electromagnetic field
\cite{Frenkel}. In the case of curved background, the interest to
the dynamics of spinning degrees of freedom has been pioneered by
Mathisson and Papapetrou \cite{Math, Papa}. For a summary of studies
before 1968, we cite the book \cite{Corben}. The self-consistent
system of equations that describes the motion of translational and
internal spinning particle degrees of freedom in the general
external electromagnetic and/or gravitational field in space-time
dimension $d=4$ has been proposed in \cite{Univ}. This model has
been re-derived is a slightly different setting in
\cite{Deriglazov14}. The recent results can be found in papers
\cite{Deriglazov, Obukhov, CQG} and references therein. The problem
of inclusion of interaction between the charged particle and
external electromagnetic and/or gravitational field exists in lower
and higher dimensions \cite{GKL, LSS1}. For particular applications
of the spinning particle concept, we refer articles \cite{A1, A2,
A3, A4, A5}.

The current models with an electromagnetic or gravitational coupling
have one common feature: the spinning particle is considered as a
point object in space-time with the internal degrees of freedom.
Depending on the parametrization of internal space coordinates
vector, tensor, spinor and twistor models of spinning particles are
distinguished \cite{Fryd}. The classical dynamics of free spinning
particle is characterized by its path in Minkowski space and the
trajectory in internal space. In all the instances, the space-time
classical trajectory is preserved by the gauge symmetries of the
theory. The introduction of internal space is essential in all the
models of point particle, because the space-time trajectory does not
contain exhaustive information about the motion of model. For
example, the space-time classical trajectory of free particle is
always a straight line, while the spin vector path in the internal
space is a point. The rectilinear motion describes the dynamics of
translational degrees of freedom, while the internal space path
makes the same job for spin. Once the space-time trajectory of the
particle is of interest, the information is lost about the motion of
spin.

In the article \cite{KLS}, a geometrical model of massive spinning
particle has been proposed. The quantization of theory corresponds
to the irreducible representation of the Poincare group with
prescribed values of mass and spin. The classical trajectory of
spinning particle is a curve on the $2d$ circular cylinder with a
time-like axis. The motion of the particle is characterized by small
irregular fluctuations of the path, whose value is determined by
spin (in this particular model). These fluctuations has been
explained by the zitterbewegung phenomenon. In the model of point
particle, there is no zitterbewegung. The phenomenon is critical for
description of spinning degrees of freedom in terms of space-time
path: the direction of path fluctiations determines the value of
spin angular momentum. The possibility of pure geometrical
description of spinning particle dynamics has been unnoticed for the
long time, even though the list of geometrical models has been
significantly extended. The spinning particle theories with a
cylindrical classical trajectories has been re-derived in
\cite{Star,Rempel}. In the theory \cite{Ners1}, the classical
trajectories are null-like helices, being isotropic paths on the
circular cylinder. For geometrical models of spinning particle in
space-time dimension three, we refer \cite{GKL, Ners, Ners3}. As for
massless particles, their path are known to lie on hyperplanes
\cite{Horv}.

A general geometric method of spinning particle dynamics description
is developed in \cite{WS}. In this work, it has been observed that
the positions of irreducible spinning particle lie on a cylindrical
surface in Minkowski space. For massive particle the surface is a
toroidal cylinder $\mathbb{R}\times \mathbb{T}^d$, where $0\leq
d\leq[(n-1)/2]$. Here, the square brackets denote the integral part
of number. The space-time submanifold that contains all the particle
positions was termed a \emph{world sheet}. In the space-time
dimension $d=3,4$, the world sheets are circular cylinders with the
time-like axis. The cylinder radius is the model parameter. The
direction of cylinder axis is determined by the momentum. The
position of the cylinder in space-time is determined by the total
angular momentum. The classical paths of the particle are general
cylindrical lines. In terms of particle positions, the equations of
motion involve higher derivatives, and they are non-Lagrangian. Once
appropriate set of auxiliary variables is introduced, the equations
of motion reproduce the previously mentioned geometrical models, see
\cite{WS} for example. The geometric description of spinning
particle dynamics by means of the world sheet concept relies uses
the conservation law of momentum and total angular momentum, so it
can be applied in the case of the free particle in its current form.

In the present work, we address an issue of description charged
spinning particle motion in the external electromagnetic field by
means of the world sheet concept. We consider the class of models
such that demonstrate the zitterbewegung already at the free level.
An inconsistency is observed for the minimal electromagnetic
coupling in these theories unless the external field is unform
\cite{KLS,GKL,Bratek}. Our main idea is that the projection of the
classical trajectory onto the cylinder axis is a special point with
a reduced gauge symmetry. Identifying this point (later termed the
center of mass point) with the classical position of point particle
from the works \cite{Univ, Deriglazov14}, we propose a variational
principle for cylindrical curves that admits inclusion of consistent
couplings with general (not necessarily uniform) electromagnetic
field. The class of gauge equivalence of classical trajectories of
the model forms a cylindrical surface whose axis is given by path of
point particle. The position of this surface in space-time is
determined by initial values of momentum and total angular momentum,
being gauge-invariant physical observables of the model. This result
means that the world sheet concept can be applied for description of
spinning particle dynamics at the interacting level.

We mostly consider the simplest model in the class, a massive
particle with nonzero spin that moves in three-dimensional
space-time. At the free level, the world sheet is a circular
hypercylinder with a time-like axis. The variational principle for
cylindrical curves is given by the action functional of paper
\cite{Deriglazov14}, where the classical position of the particle is
replaced by the center of mass coordinate.\footnote{The action
principle for the irreducible spinning particle travelling
cylindrical path in three-dimensional Minkowski space first proposed
in \cite{GKL} with spinor parametrization of internal space.} An
important difference is that the dynamical variables are
three-vectors, not four-vectors. The model parameters are particle's
mass, particle's spin, and the cylinder radius. The latter quantity
is an accessory parameter that distinguishes different spinning
particle models with one and the same value of mass and spin.  The
inclusion of interactions with electromagnetic field along the lines
of \cite{Univ, Deriglazov14} preserves the consistency of the model
because the theories are connected by the canonical transformation
of phase-space variables.  This brings us to the class of consistent
interactions between the electromagnetic field and spinning particle
that moves a cylindrical path.

The paper is organized as follows. In Section 2, we discuss the
classical dynamics of spinning particle model, whose quantization
corresponds to the irreducible representation of the Poincare group.
We describe geometry of the world sheet of free spinning particle,
and derive the equations of motion of the theory. In Section 3, we
propose a variational principle for cylindrical curves and construct
the class of consistent interactions with the electromagnetic field.
The model demonstrates the zitterbewegung at the free and
interacting levels. In Section 4, we derive the equations of motion
for the cylinder axis and propose a constructive procedure of world
sheet description. The classical trajectories of interacting model
are general curves on the world sheet. In Section 4, we consider the
spinning particle motion in the uniform electric field. If the
particle initial velocity is orthogonal to the electric vector, the
world sheet is shown to be a pseudotorus, whose position is
determined by the initial values of momentum and total angular
momentum. The conclusion summarizes results.

\section{Classical dynamics of irreducible spinning particle}

We consider a massive spinning particle that travels in $3d$
Minkowski space. The particle state is determined by the values of
the momentum $p$ and total angular momentum $J$, being subjected to
the mass shell and spin shell conditions
\begin{equation}\label{pJ-free}
    (p,p)+m^2=0\,,\qquad (p,J)-ms=0\,.
\end{equation}
Here, $m$ is the mass, and $s$ is spin. We assume that $m>0$. The
round brackets denote the scalar product with respect to the
Minkowski metric, $(a,b)=a_\mu b^\mu$.\footnote{We use a mostly
positive signature of the Minkowski metric throughout the paper. All
the tensor indices are raised and lowered by the metric.} The spin
angular momentum $\mathcal{M}$ is defined by the following relation:
\begin{equation}\label{S-free}
    \mathcal{M}=J-[x,p]\,,\qquad [x,p]=\epsilon_{\mu\nu\rho}x^\mu
    p^\nu dx^\rho\,.
\end{equation}
The quantity $\epsilon_{\mu\nu\rho}$ denotes the $3d$ Levi-Civita
symbol, being antisymmetric in all indices. We use the convention
$\epsilon_{012}=1.$ The symbol $x$ denotes the particle coordinate.
The article \cite{WS} tells that $\mathcal{M}$ is normalized in
every irreducible spinning particle theory,
\begin{equation}\label{M-free}
    (\mathcal{M},\mathcal{M})=m^2r^2-s^2\,,
\end{equation}
with $r$ being a real nonnegative constant with the length
dimension. The value of this constant is determined by the
representation. The possible models of spinning particles are
characterized by the values of three independent parameters: $m$,
$s$, and $r$. The mass $m$ and spin $s$ determine the parameters of
the spinning particle. The constant $r$ distinguishes particle
theories with one and the same value of mass and spin.

Relations (\ref{pJ-free}), (\ref{S-free}) and (\ref{M-free}) imply
that the possible spinning particle positions lie on a hypersurface
in Minkowski space. The equation of hypersurface eventually reads
\begin{equation}\label{ws}
    (x-y(0))^2+(n(0),x)^2-r^2=0\,,
\end{equation}
where $x$ is the particle coordinate, and the notation is used,
\begin{equation}\label{pJ-ny}
    y(0)=\Big[\frac{p}{m},\frac{J}{m}\Big],\qquad
    n(0)=\frac{p}{m}\,\qquad\Leftrightarrow\qquad p=mn(0)\,,\qquad
    J=m[y(0),n(0)]-sn(0)\,.
\end{equation}
By construction, the vector $n(0)$ is normalized and orthogonal to
$y(0)$,
\begin{equation}\label{ny}
    (n(0),n(0))+1=0\,,\qquad (n(0),y(0))=0\,.
\end{equation}
Equation (\ref{ws}) determines a hypercylinder in Minkowski space
with the time-like axis. The vector $n(0)$ is tangent to the
cylinder axis. The vector $y(0)$ connects the cylinder axis and
origin by the shortest path. The quantity $r$ defines the radius of
hypercylinder. The cylinder of zero radius is a straight line with
the time-like tangent vector. Relation (\ref{ws}) is a unique
condition imposed onto the particle coordinate $x$, so the
hypercylinder includes all the possible spinning particle positions.
This means that the circular hypercylinder with the time-like axis
represents a set of massive spinning particle positions in three
dimensional Minkowski space.

The conditions (\ref{pJ-free}), (\ref{M-free}) and their consequence
(\ref{ws}) determine the classical dynamics of model. The classical
trajectory represents a set of consecutive positions that are
occupied by the spinning particle. Once the particle positions lie
on a hypercylinder, the classical paths are cylindrical curves. The
article \cite{Starostin} tells us that the cylindrical curves are
determined by the following system of ordinary differential
equations:
\begin{equation}\label{csc}
    \frac{d}{d\uptau}\bigg(\frac{\dot{y}}{\sqrt{-\dot{y}^2}}\bigg)=0\,,\qquad
    (\dot{y},\omega)=0\,,\qquad (\omega,\omega)-r^2=0\,.
\end{equation}
Here, $y$ denotes the coordinate of projection spinning particle
current position onto the axis of hypercylinder. The difference
vector $\omega$ connects the particle position and cylinder axis by
the shortest path. The classical position of the particle are
determined by the rule
\begin{equation}\label{d-vec}
    x=y+\omega.
\end{equation}
The quantity $\uptau$ denotes the proper time. The dot denotes the
derivative by $\uptau$. The dynamical variables in the system
(\ref{csc}) are $y$ and $\omega$. Relation (\ref{d-vec}) allows us
to chose two alternative sets of independent variables: $x$ and
$\omega$, $x$ and $y$. In what follows, the variable $y$ is called
the coordinate of mass center. We justify this terminology by
observing that the classical trajectory of free particle mass center
is always a line.

The system (\ref{csc}) determines the class of cylindrical curves
indeed. The first equation in the set (\ref{csc}) tells us that the
curve $y(\uptau)$ is a straight line with time-like tangent vector.
Each line is determined by two ingredients: the direction of its
tangent vector and position of initial point. If $\uptau=0$ is the
initial moment, the normalized time-like vector $n(0)$ (\ref{ny}) is
tangent to the line. The initial value $y(0)$ of the variable $y$ is
the other Cauchy data. As the theory is reparametrization-invariant,
there no selected initial point on the line. The second condition
(\ref{ny}) fixes a particular choice for initial position of mass
center $y(0)$. This choice has no physical meaning because it
corresponds to the particular gauge fixing. Relations (\ref{pJ-ny})
tell us that the position cylinder axis contains all the valuable
information about the dynamics of massive spinning particle in $3d$
space. Second and third conditions in the set (\ref{csc}) determine
the relative position of the particle and cylinder axis. They tell
us that the particle lies on circle of radius $r$ with the center in
point $y$. The vector $n(0)$ is orthogonal to the plane, which
contains a circle. With the proper time being, the vector $y$ runs
over the line, and $x$ moves along the general cylindrical path.
This means that the cylindrical lines are solutions to equations
(\ref{csc}). The classical trajectories of spinning particle
described by these equations.

Let us now explain the concept of spinning particle world sheet.
Equation (\ref{pJ-ny}) connects the parameters of hypercylinder with
the momentum and total angular of particle. Since this
correspondence is a bijection, there is one-to-one relationship
between the elements of hypercylinder set (\ref{ws}), (\ref{ny}) and
particle states. This one-to-one correspondence is organized as the
follows. Each particle state is determined by the values of momentum
and total angular momentum, being subjected to the mass shell and
spin shell conditions (\ref{pJ-free}). The world sheet (\ref{ws})
represents the set of particle positions in space-time, the particle
state is known. On the other hand, each hypercylinder of radius $r$
determines the state of spinning particle with the mass $m$ and spin
$s$ by the rule (\ref{pJ-ny}). The cylinder radius determines the
norm of spin angular momentum by the rule (\ref{M-free}). The
cylindrical surface that contains all the possible positions of
spinning particle was termed a world sheet in \cite{WS}. The world
sheet concept reduces the problem of description of spinning
particle dynamics to the task of classification of general curves on
certain set of (hyper)surfaces in Minkowski space. In the case of
massive spinning particles in $3d$ space-time, the circular
hypercylinders with the time-like axis are of interest. The
classical trajectories of the model are cylindrical curves, being
described by equation (\ref{csc}). The relation (\ref{pJ-ny})
connects the initial data for equations (\ref{csc}) and values of
momentum and total angular momentum of the spinning particle.

The description of spinning particle dynamics by means of the world
sheet formalism has two important subtleties. First, the world sheet
dimension depends on the values of model parameters. For example,
the world sheet of spinning particle (\ref{ws}) is hypersurface
(codimension 1) for $r>0$. For $r=0$, it is a line (codimension 2).
The example of the work \cite{WS4} tells us that the world sheets of
maximal dimension determine spinning particle state in a unique way.
For the world sheets of submaximal dimension, the information about
the spin state is lost. The case $r=0$ in $3d$ space-time is not
representative because spin has no physical degrees of freedom in
this model. The theories with zero and nonzero $r$ have slightly
different properties. For $r>0$, the classical spinning particle
trajectory can deviate from the rectilinear path, and the
zittebewegung phenomenon is observed. For $r>0$, the particle path
is always a line. This is a model without zitterbewegung at the free
level. In the current article, we consider the theories with nonzero
cylinder radius. As it has been mentioned above, the models with the
zitterbewegung at the free level have obstructions for inclusion of
consistent interactions with general electromagnetic and
gravitational field. Once the world sheets of interest, the
consistent couplings between the changed spinning particle and
external electromagnetic field has to be constructed in the model
with zitterbewegung.

Our second remark is that the spinning particle classical trajectory
is always a world line, not a world sheet. The world line position
in space-time determines a particle state unambiguously if the curve
lies on a unique world sheet. If the world line lies on the
intersection of several world sheets, multiple states can be
associated with one and the same curve. In the article \cite{WS} the
curves that lie on a unique world sheet has been termed typical. The
paths that lie on several world sheets has been called atypical. In
the case of $3d$ Minkowski space, the straight lines with the
time-like tangent vector provide an example of atypical curves. A
single line belongs to the infinite number of circular
hypercylinders with one and the same direction of axis. In what
follows, we ignore the rectilinear path. The other atypical curves
are closed loops that lie on intersection of pairs of cylinders. To
exclude loops the casuality condition $\dot{x}^0>0\,$ is imposed
onto the classical paths of spinning particle. The causal classical
trajectories, which are not straight lines, are spinning particle
path of interest. These curves determine the state of spinning
particle in a unique way.

The described above exposition of the world sheet concept
essentially uses the conservation law of momentum and total angular
momentum. This is true for free particle. In the next sections, we
address two following aspects of the spinning particle dynamics. We
construct the system of ordinary differential equations that
describe the motion of charged spinning particle with general value
of invariants $m$, $s$, $r$ (\ref{pJ-free}), (\ref{M-free}) in the
electromagnetic field. We demonstrate that the class of gauge
equivalence of spinning particle trajectories forms a cylindrical
hypersurface in Minkowski space irrespectively to specifics of the
field configuration. The quantity $r$ determines the radius of
cylinder. The shape of symmetry axis of hypersurface is determined
by the field configuration. The position of the world sheet is
determined by the initial values of momentum and total angular
momentum. The results of our study show that the concept of world
sheet of spinning particle can be used for the description of
dynamics even in the presence of external electromagnetic field. The
classical trajectory of the particle, which lies on the world sheet,
demonstrates the zitterbewegung even if the external electromagnetic
is nonzero.

\section{Variational principle and inclusion of interaction}

In this section, we derive a variational principle for the equations
(\ref{csc}) for cylindrical curves and propose the procedure of
inclusion of interaction between the charged spinning particle and
electromagnetic field. The section is organized as follows. At
first, we construct the Hamiltonian variational principle for
equations (\ref{csc}). The vector parametrization of internal space
is used. The phase space-variables are subjected to constraints. We
explicitly verify that the Noether charges, being associated with
the space-time translations and Lorentz rotations, meet the mass
shell and spin shell conditions (\ref{pJ-free}). This ensures that
the canonical quantization of classical theory corresponds to the
irreducible representation of the Poincare group with the mass $m$
and spin $s$. The gauge symmetries of the model are identified. It
shown that they connect all the paths on one and the same cylinder.
Then, we propose the procedure of inclusion of interaction between
the particle and electromagnetic field. The interaction preserves
both gauge symmetries of the model, even though the zitterbewegung
is observed at the free level.

We chose the direct product $\mathbb{R}^{2,1}\times
\mathbb{R}^{2,1}$ as the configuration space of the spinning
particle. The first factor is interpreted as Minkowski space, while
the second one is attributed to the internal space of spinning
particle. In the old-school terminology, such structure of
configuration space corresponds to the vector model of spinning
particle \cite{Fryd}. Three different choices for the dynamical
variables are admissible. In the position representation, the set of
generalized coordinates includes the classical particle position $x$
and difference vector $\omega$. The position representation
corresponds to the most common choice for the dynamical variables in
spinning particle theories, see \cite{GKL} for example. In the
center of mass representation, the set of generalized coordinates
includes the classical position of mass center $y$ and difference
vector $\omega$. If $x$ and $y$ are identified, the possibility
corresponds to the spinning particle model with the world sheets of
dimension 1. The classical trajectories of the point particle are
straight lines. No zitterbewegung is observed at the free level. The
procedure of interaction inclusion is known only in center of mass
representation \cite{Univ, Deriglazov14}. The third option for the
generalized coordinates is $x$ and $y$. This corresponds to some
sort of bivector model of spinning particle \cite{Rempel}. The
relation (\ref{d-vec}) connects various representations of dynamics.
We mostly use the center of mass representation is the current
section because it leads to the simplest form of dynamical equations
for the spinning particle classical trajectory. We return to the
position representation in the process of classification of world
sheets in Section 4. We do not use the bivector representation for
the particle dynamics in the article.

We proceed with the introduction of canonical momenta $p$, $\pi$ for
the generalized coordinates $y,\omega$ in the center of mass
representation. In other representations, the canonical momenta are
introduced by canonical transformation that is generated by the
change of generalized coordinates (\ref{d-vec}). We do not introduce
for momenta in other representations of dynamics because they are
not relevant in this section. The following constraints are imposed
onto the phase-space variables:
\begin{equation}\label{S-constr}\begin{array}{c}\displaystyle
    \Theta_e=(p,p)+m^2\approx0\,,\qquad \Theta_2=(\omega,\omega)-r^2\approx0\,,\qquad
    \Theta_3=(\pi,\pi)-r^{-2}s^2\approx0\,,\\[5mm]\displaystyle
    \Theta_5=(p,\omega)\approx0\,,\qquad \Theta_4=(\omega,\pi)\approx0\,,\qquad
    \Theta_6=(\omega,\pi)\approx0\,.
\end{array}\end{equation}
The sign $\approx$ means equality on the mass shell. Relations
(\ref{S-constr}) have a clear origin. The first equation tells us
that the linear momentum meets the mass shell condition
(\ref{pJ-free}). Relations $\Theta_2,\Theta_4\approx0$ imply the the
difference vector $\omega$ is normalized and orthogonal to momentum
$p$. These conditions determine relative positions of the particle
and its mass center. Thee remaining relations express the vector
$\pi$ in terms of the dynamical variables $p,\omega$. They mean that
$\pi$ is an auxiliary variable, having no independent dynamical
degrees of freedom. The particular form for the constraints
$\Theta_3,\Theta_5,\Theta_6\approx0$ is chosen to satisfy the spin
shell condition (\ref{pJ-free}). The Hamiltonian action principle
for the spinning particle model reads
\begin{equation}\label{S-ms}\begin{array}{c}\displaystyle
    S=\int
    \bigg(p\dot{x}+\pi\dot{n}-1/2(e\Theta_e+\lambda_2\Theta_2+\lambda_3\Theta_3)-
    \lambda_4\Theta_4-\lambda_5\Theta_5-\lambda_6\Theta_6\bigg)d\uptau\,,
\end{array}\end{equation}
The dynamical variables are the generalized coordinates $y,\omega$,
canonically conjugated momenta $p,\pi$, and Lagrange multipliers
$e,\lambda_\alpha,\alpha=2,\ldots,6$. The action functional
(\ref{S-ms}) has been proposed  for a massive spinning particle
traveling in four-dimensional Minkowski space in the paper
\cite{Deriglazov14}. In that work, the generalized coordinate $y$
has been associated with the particle position, not the coordinate
of the center of mass point. In our model, the center of mass
coordinate follows a rectilinear path, while the classical
trajectory of the particle lies on a circular hypercylinder with the
time-like axis of radius $r$. Once the particle path is not a
straight line, the particle theory (\ref{S-ms}) demonstrates
zitterbewegung already at the free level.

The model (\ref{S-ms}) describes an irreducible spinning particle
indeed. The momentum and total angular momentum are the conserved
charges, being associated with the invariance of the theory with
respect to the space-time translations and Lorentz rotations,
\begin{equation}\label{pJ-exp}
    \phantom{\frac12}\mathcal{P}=p,\qquad \mathcal{J}=[y,p]+[\omega,\pi]\,.\phantom{\frac12}
\end{equation}
The  constraints (\ref{S-constr}) imply the mass shell and spin
shell conditions for the quantities $\mathcal{P}$, $\mathcal{J}$,
\begin{equation}\label{ms-shell}
    \phantom{\frac12}(\mathcal{P},\mathcal{P})+m^2=0,\qquad (\mathcal{P},\mathcal{J})-ms=0\,.\phantom{\frac12}
\end{equation}
The spin angular momentum reads
\begin{equation}\label{M-def}
    \phantom{\frac12}\mathcal{M}=[\omega,\pi+p]\,.\phantom{\frac12}
\end{equation}
By construction, the vector $\mathcal{M}$ is normalized in the sense
(\ref{M-free}). Relations (\ref{ms-shell}), (\ref{M-def}) ensure
that the theory (\ref{S-ms}) describes an irreducible spinning
particle theory with general values of invariants $m$, $s$, $r$. The
relationship between equations (\ref{S-constr}) and (\ref{ms-shell})
has an important subtlety. The constraints (\ref{S-constr}) imply
that the vectors $p$, $[\omega,\pi]$ are normalized and collinear,
\begin{equation}\label{chir}
    \frac{p}{m}=k\frac{[\omega,\pi]}{m|s|}\,,\qquad k=\pm1\,.
\end{equation}
The symbol $|s|$ denotes the absolute value of $s$. The quantity $k$
can be interpreted as the chirality of the particle. For $k=1$, the
vectors $p$, $\omega$, $\pi$ form a right-handed triple. For $k=-1$,
the vectors $p$, $\omega$, $\pi$ form a left-handed triple. The spin
shell condition (\ref{ms-shell}) follows from (\ref{S-constr}),
(\ref{pJ-exp}) if the chirality spin have one and the same sign,
\begin{equation}\label{chir-1}
    \phantom{\frac12}k=\text{sgn}(s)\,,\phantom{\frac12}
\end{equation}
This means that the right-handed triple of vectors $p$, $\omega$,
$\pi$ describes a spinning particle with a positive value of spin.
The left handed triple describes a spinning particle with a negative
value of spin. Contrary to the physical intuition, the chirality is
not connected to the shape of spinning particle trajectory. As we
see in the next paragraph, the particle trajectories with positive
or negative spin may have one and the same shape.

The conditions of conservation of the constraints (\ref{S-constr})
imply that four of six Lagrange multipliers are expressed on the
mass shell,
\begin{equation}\label{}
    \phantom{\frac12}\lambda_2=s^2r^{-4}\lambda_3=0\,,\qquad \lambda_4=\lambda_5=\lambda_6=0.\phantom{\frac12}
\end{equation}
The quantities $e$ and $\lambda_3$ remain arbitrary functions of the
proper time $\uptau$. The equations of motion for the phase-space
variables of the model read
\begin{equation}\label{EoM-free-cma}
    \phantom{\frac12}\dot{y}=ep\,,\qquad \dot{\omega}=\lambda_3\pi\,,\qquad
    \dot{\pi}=-\frac{s^2}{r^4}\lambda_2\omega\,,\qquad \dot{p}=0\,.\phantom{\frac12}
\end{equation}
In the position representation, the equations of motion for the
particle coordinate $x$ and difference vector $\omega$ have the
following form:
\begin{equation}\label{EoM-free-pos}
    \phantom{\frac12}\dot{x}=ep+\lambda_3\pi\,,\qquad \dot{\omega}=\lambda_2\pi\,.\phantom{\frac12}
\end{equation}
Relations (\ref{EoM-free-cma}), (\ref{EoM-free-pos}) tell us that
classical trajectory of the mass center is a line, while vectors
$\omega,\pi$ rotate around the plane with normal $p$. The quantity
$\lambda_3$ determines the angular velocity of rotation. The sign of
$\lambda_3$ defines the direction of rotation. Since $\lambda_3$ can
be positive and negative, both the directions of rotation are
admissible. In the position representation, the dynamics of the
vector $\omega$ determines the shape of the space-time particle
trajectory. The positive direction of rotation corresponds to the
right-handed cylindrical curves. The negative direction of rotation
corresponds to the left-handed cylindrical curves. Since both
directions of rotations are admissible in every spinning particle
theory, the chirality of space-time trajectory does not related with
the value of spin. The rectilinear spinning particle trajectories
are selected by the condition $\lambda_3=0$.

Equations (\ref{EoM-free-cma}) are preserved by two gauge
symmetries. The first gauge transformation generates translations
along the center of mass trajectory,
\begin{equation}\label{gt-1-free}
    \phantom{\frac12}\delta_\xi y=p\xi,\qquad \delta_\xi \omega=\delta_\xi
    \pi=\delta_\xi p=0\,,\phantom{\frac12}
\end{equation}
where $\xi=\xi(\uptau)$ is the parameter, being arbitrary function
of proper time. The second gauge transformation generates rotations
in $\omega,\pi$ plane,
\begin{equation}\label{gt-2-free}
    \phantom{\frac12}\delta_\zeta \omega=\pi\zeta,\qquad \delta_\xi \pi=-s^2r^{-4}\omega\zeta\,,\qquad
    \delta_\zeta y=\delta_\zeta p=0\,.\phantom{\frac12}
\end{equation}
The transformation parameter is the arbitrary function of proper
time $\zeta=\zeta(\uptau)$. The gauge symmetries (\ref{gt-1-free}),
(\ref{gt-2-free}) have clear physical meaning. The translations
along the center of mass classical trajectory (\ref{gt-1-free})
generate the shifts of the particle path along the cylinder. The
gauge symmetry (\ref{gt-2-free}) generates the shifts of the
particle classical trajectory in the direction, which is orthogonal
to the cylinder axis. These shifts can be considered as
infinitesimal rotations the particle path around the axis of
cylinder. As the cylinder (\ref{ws}) is a two-dimensional surface,
we conclude that almost all classical paths on the cylinder are
connected by gauge transformations. The spinning particle admits a
special class of trajectories with a reduced gauge symmetry, being
straight lines. We exclude the rectilinear trajectories from our
consideration because they lie on infinite number of cylinders at
one and the same moment.

Let us now turn to the problem of inclusion of interactions between
the spinning particle and external electromagnetic field. In the
most general setting, the consistent interactions between gauge
fields are associated with the deformations of action functional
that preserve all the gauge symmetries and number of physical
degrees of freedom. The problem of inclusion of interaction between
the spinning particle and general electromagnetic field has been
first solved in the model with the spinor parametrization of
internal space in \cite{Univ}. The same problem was solved in the
vector model in \cite{Deriglazov14}. In both the theories the $d=4$
model with zero cylinder radius is considered. The classical
trajectory of the spinning particle does not demonstrate
zitterbewegung at the free level. Below, we apply the procedure of
the articles \cite{Univ, Deriglazov14} for construction of
consistent interactions between the charged spinning particle and
electromagnetic field in the center of mass representation in the
model with the general value cylinder radius $r$. In this case, the
relationship (\ref{d-vec}) between the position of center of mass
representation guarantees that the classical trajectory of spinning
particle demonstrate the zitterbewegung in the presence of external
electromagnetic field. No assumptions about the configuration of the
electromagnetic field are done, it can be general. To our knowledge,
purposed model is the first spinning particle theory with
zitterbewegung which admits consistent coupling with the
electromagnetic field of general configuration. This result
demonstrates that presence of zitterbewegung at the free level is
not an obstruction for construction of consistent couplings.

At first, we explain the general idea of interaction constriction.
Once the electromagnetic field is in question, the extended momentum
is a relevant object. In the current article, the extended momenta
is defined by the rule
\begin{equation}\label{ext-P}
    \phantom{\frac12}P_\mu=p_\mu+A_\mu(y)\,.\phantom{\frac12}
\end{equation}
Here, the particle charge is set to $-1$, and $A_\mu$ denotes the
vector potential of the electromagnetic field. In the formula
(\ref{ext-P}), the vector potential is taken at the mass center
point $y$. In the standard definition of the extended momentum, the
vector potential is taken at the current particle position $x$. The
dependence of the vector potential on the center of mass position is
essential for consistency of interaction. The particle coordinate
$x$ transforms by two gauge symmetries: translations along the
cylinder axis and rotations around it. Once the moment of time is
fixed, the set of particle positions is given by the section of
cylinder, being circle or ellipse. All the points of the section
represent equivalent particle positions, which are indistinguishable
from the physical viewpoint. The vector potential describing the
interaction between the charged particle and electromagnetic field
must be gauge-invariant. This is possible if it depends on the
coordinates of the mass center, being the special point with a
reduced gauge symmetry. The dependence of the vector potential on
the mass center coordinate has no alternative. The analysis of paper
\cite{GKL} tells us that the coupling with the standard definition
of extended momentum is consistent in the anyon model if the
electromagnetic field strength meets equation of Chern-Simons type.
The components of extended momentum do not commute. The Poisson
bracket read
\begin{equation}\label{}
    \phantom{\frac12}\{P_\mu,P_\nu\}=\epsilon_{\mu\nu\rho}F^\rho(y)\,,\phantom{\frac12}
\end{equation}
where $F_\mu(y)=\epsilon_{\mu\nu\rho}\partial^\nu A^\rho(y)$ is the
vector of electromagnetic field strength in the center of mass
position $y$ (the partial derivative acts on $y$).

We construct the set of constraints at the interacting level as
follows: the canonical momenta $p_\mu$ are replaced by the extended
momenta $P_\mu$, and the non-minimal coupling term is included into
the mass shell condition. By definition, we put
\begin{equation}\label{constr-pos-rep-int}\begin{array}{c}\displaystyle
    \widetilde{\Theta}_e=(P,P)-g(M,F)+m^2\approx0\,,\quad \widetilde{\Theta}_2=(\pi,\pi)-s^2/r^2\approx0\,,\quad
    \widetilde{\Theta}_3=(\omega,\omega)-r^2\approx0\,,\\[5mm]\displaystyle
    \widetilde{\Theta}_4=(P,\omega)\approx0\,,\qquad \widetilde{\Theta}_5=(P,\pi)\approx0\,,\qquad
    \widetilde{\Theta}_6=(\pi,\omega)\approx0\,.
\end{array}\end{equation}
Besides the mass and spin, the interaction involves an additional
parameter $g$, being the gyromagnetic ratio or $g$-factor. If the
particle has no anomalous magnetic dipole moment, $g=2$. The vector
\begin{equation}\label{M-Frenkel}
    M=[\omega,\pi]
\end{equation}
denotes the Frenkel angular momentum. Relations (\ref{pJ-free}),
(\ref{S-free}) imply that the quantity $M$ coincides with the total
angular momentum of the particle in the rest system of its mass
center. The Frenkel angular momentum is invariant with respect to
the gauge transformations (\ref{gt-1-free}), (\ref{gt-2-free}). This
provides another interpretation for $M$: (\ref{M-Frenkel}) it is a
gauge-invariant part of the total angular momentum $J$. The Frenkel
angular momentum and spin angular momentum are equal if the cylinder
radius is zero, $r=0$. In the spinor parametrization of internal
space, the system of constraints has been first proposed
(\ref{constr-pos-rep-int}) in \cite{Univ}. The vector
parametrization of internal space has been used in
\cite{Deriglazov14}. In both mentioned above papers, the special
model with absence of zitterbewegung at free level has been
considered (this model corresponds to zero cylinder radius limit in
the Lagrangian (\ref{S-ms})). In the case of theory
(\ref{constr-pos-rep-int}), the cylinder radius is arbitrary
positive constant. The interaction (\ref{constr-pos-rep-int}) is
consistent because it is connected with that of the paper
\cite{Univ, Deriglazov14} by the point canonical transformation,
which is induced by the coordinate change (\ref{d-vec}). In next
paragraph, we demonstrate that both the gauge symmetries of free
theory (\ref{S-ms}) are preserved by coupling.

The variational principle for the interacting model reads
\begin{equation}\label{S-ms-int}
    S=\int
    (p\dot{y}+\pi\dot{\omega}-1/2(e\widetilde{\Theta}_e+\lambda_2\widetilde{\Theta}_2+
    \lambda_3\widetilde{\Theta}_3)-
    \lambda_4\widetilde{\Theta}_4-\lambda_5\widetilde{\Theta}_5-\lambda_6\widetilde{\Theta}_6)d\uptau\,.
\end{equation}
The dynamical variables are generalized coordinates $y,\omega$,
momenta $p,\pi$, and Lagrange multipliers
$e,\lambda_\alpha,\alpha=2,\ldots,6$. The condition of conservation
of constraints (\ref{constr-pos-rep-int}) allows us to determine
four Lagrange multipliers of six,
\begin{equation}\label{}\begin{array}{c}\displaystyle
    \lambda_6=0\,,\qquad\lambda_2=\frac{s^2}{r^4}\lambda_3\,,
    \qquad \lambda_4=\frac{1}{2}\frac{(g-2)(\pi,P,F)-g(\pi,\partial)(M,F)}{m^2-(g+1)(M,F)}\,,
    \\[5mm]\displaystyle
    \lambda_5=-\frac{1}{2}\frac{(g-2)(\omega,P,F)-g(\omega,\partial)(M,F)}{m^2-(g+1)(M,F)}\,.
\end{array}\end{equation}
The quantities $e$ and $\lambda_3$ remain arbitrary functions of
proper time. The following Lagrange equations follows from the least
action principle for the functional (\ref{S-ms-int}):
\begin{equation}\label{EoM-y}
     \phantom{\sum_1^2}\frac{dy}{d\uptau}=e\Big\{P+\frac{1}{2}\frac{(g-2)(\pi,P,F)-g(\pi,\partial)(M,F)}{m^2-(g+1)(M,F)}\omega
    -(\omega\leftrightarrow\pi)\Big\}\,, \phantom{\sum_1^2}
\end{equation}
\begin{equation}\label{EoM-P}
     \phantom{\sum_1^2}\frac{dP}{d\uptau}=e\Big\{\frac{g}{2}\partial(M,F)+
    \frac{1}{2}\Big[\frac{(g-2)(\pi,P,F)-g(\pi,\partial)(M,F)}{m^2-(g+1)(M,F)}\omega
    -(\omega\leftrightarrow\pi),F\Big]\Big\}\,, \phantom{\sum_1^2}
\end{equation}
\begin{equation}\label{EoM-pi}
    \phantom{\sum_1^2}\frac{d\pi}{d\uptau}=-\frac{s^2}{r^4}\lambda_3
    \omega-e\frac{1}{2}\frac{(g-2)(\pi,P,F)-g(\pi,\partial)(M,F)}{m^2-(g+1)(M,F)}P\,, \phantom{\sum_1^2}
\end{equation}
\begin{equation}\label{EoM-om}
     \phantom{\sum_1^2}\frac{d\omega}{d\uptau}=\lambda_3\pi-e\frac{1}{2}\frac{(g-2)(\omega,P,F)
    -g(\omega,\partial)(M,F)}{m^2-(g+1)(M,F)}P\,. \phantom{\sum_1^2}
\end{equation}
The first equation in this system expresses the spinning particle
extended momentum in terms of center of mass coordinate and its
velocity. If the electromagnetic field strength is nonzero, the
velocity $\dot{y}$ and extended momentum $P$ are not collinear.
Equation (\ref{EoM-P}) impose the second order differential equation
for spinning particle classical trajectory, $y=y(\uptau)$. Relations
(\ref{EoM-pi}), (\ref{EoM-om}) determine the time evolution of
internal space variables $\omega$, $\pi$. This evolution is not
casual because $\omega$ has one independent component modulo
constraints (\ref{constr-pos-rep-int}). The time derivative of this
component is determined by the Lagrange multiplier $\lambda_3$,
being arbitrary function of proper time.

The system of equations (\ref{EoM-y}), (\ref{EoM-P}),
(\ref{EoM-pi}), (\ref{EoM-om}) is preserved by two gauge symmetries
that generalize free infinitesimal transformations
(\ref{gt-1-free}), (\ref{gt-2-free}),
\begin{equation}\label{gt-1-1}
     \phantom{\sum_1^2}\delta_\xi y=\Big\{P+\frac{1}{2}\frac{(g-2)(\pi,P,F)-
     g(\pi,\partial)(M,F)}{m^2-(g+1)(M,F)}\omega
     -(\omega\leftrightarrow\pi)\Big\}\xi\,, \phantom{\sum_1^2}
\end{equation}
\begin{equation}\label{gt-1-2}
     \phantom{\sum_1^2}\delta_\xi P=\Big\{\frac{g}{2}\partial(M,F)+
     \frac12\Big[\frac{(g-2)(\pi,P,F)-g(\pi,\partial)(M,F)}{m^2-(g+1)(M,F)}\omega
     -(\omega\leftrightarrow\pi),F\Big]\Big\}\xi\,, \phantom{\sum_1^2}
\end{equation}
\begin{equation}\label{gt-1-3}
     \phantom{\sum_1^2}\delta_\xi\pi=-\frac{1}{2}\frac{(g-2)(\pi,P,F)-g(\pi,\partial)(M,F)}
     {m^2-(g+1)(M,F)}P\xi\,, \phantom{\sum_1^2}
\end{equation}
\begin{equation}\label{gt-1-4}
     \phantom{\sum_1^2}\delta_\xi \omega=-\frac12\frac{(g-2)(\omega,P,F)
     -g(\omega,\partial)(M,F)}{m^2-(g+1)(M,F)}P\xi\,, \phantom{\sum_1^2}
\end{equation}
and
\begin{equation}\label{gt-2}
    \phantom{\frac12}\delta_\zeta y=\delta_\zeta P=0\,,\qquad \delta_\zeta \omega=\pi\zeta\,,\qquad \delta_\zeta \pi=-r^{-4}s^2\zeta\,.\phantom{\frac12}
\end{equation}
The gauge transformation parameters are arbitrary functions of
proper time $\xi,\zeta$. The geometric meaning of the
transformations is the same as in the free case. The first gauge
symmetry generates shifts of particle position along the center of
mass trajectory. The second gauge generates rotations around the
current center of mass position. The extended momenta $P$ determines
the normal to the plane of rotation.

\section{Motion of mass center and world sheet}

In this section, we consider the general properties of spinning
particle dynamics in external electromagnetic field. At first, we
observe that the spin has no physical degrees of freedom in $3d$
space-time. This allows us to construct a closed system of equations
that describes the time evolution of center of mass coordinate,
extended momentum, and Frenkel angular momentum. Then, the Frenkel
angular momentum is expressed from this system, and we get equations
of motion for the mass center trajectory in terms of variables $y$
and $P$. The exclusion of the extended momentum from this system
(which we do not make here) leads the the second-order differential
equation for center of mass path of the spinning particle. The
existence of this system is a consequence of small space-time
dimension. In the space-time dimension $d=4$, where spin has two
physical polarizations, the center of mass trajectory does not
determine the particle dynamics. Then, we consider the spinning
particle dynamics in the position representation. We show that the
spinning particle trajectories lie on an cylindrical hyper surface
in Minkowski space and provide a constructive procedure of world
sheet construction.

We begin our consideration from the construction of a closed system
of evolutionary equations for three variables: $y$, $P$, and $M$.
The double vector product formula,\footnote{We note that the signs
in the right hand side of relation are sensitive to the signature of
the metric.}
\begin{equation}\label{}
    \phantom{\frac12}[a,[b,c]]=c(a,b)-b(a,c),\phantom{\frac12}
\end{equation}
brings equations (\ref{EoM-y}), (\ref{EoM-P}) to the desired form
\begin{equation}\label{EoM-yP}
     \frac{1}{e}\frac{dy}{d\uptau}=
     P+\frac{1}{2}\bigg[M,\frac{(g-2)[P,F]-g\partial(M,F)}{m^2-(g+1)(M,F)}\bigg]\,,\qquad
     \frac{1}{e}\frac{dP}{d\uptau}=\frac{g}{2}\partial(M,F)+
     \bigg[\frac{1}{e}\frac{dy}{d\uptau},\,F\bigg]\,.
\end{equation}
Relations (\ref{EoM-pi}), (\ref{EoM-om}) determine the motion of the
Frenkel angular momentum,
\begin{equation}\label{EoM-Mc}
    \phantom{\frac12}\frac{1}{e}\frac{dM}{d\uptau}=
    \bigg[P,\frac{1}{e}\frac{dP}{d\uptau}-\frac{g}{2}\partial(M,F)\bigg]\,.\phantom{\frac12}
\end{equation}
The dynamical equations (\ref{EoM-yP}), (\ref{EoM-Mc}) are
supplemented by the constraints
\begin{equation}\label{PMc-constr}
    \phantom{\frac12}(P,P)-g(M,F)+m^2\approx0\,,\qquad (M,M)+s^2\approx0\,.\phantom{\frac12}
\end{equation}
\begin{equation}\label{M-P}
    [P,M]=0\,.
\end{equation}
The conditions (\ref{PMc-constr}), (\ref{M-P}) are consequences of
constraints (\ref{constr-pos-rep-int}) and definition of the
variable $M$ (\ref{M-Frenkel}). Equations (\ref{PMc-constr}) express
the mass shell and spin shell conditions for $P$,$M$. Relation
(\ref{M-P}) implies that vectors of extended momentum and Frenkel
angular momentum are collinear. Relations (\ref{EoM-yP}),
(\ref{EoM-Mc}) represent a three-dimensional analog of the Frenkel
equations, which has been derived in \cite{Deriglazov14}. The
spacial feature of three-dimensional case is that the Frenkel
angular momentum $M$ can be expressed algebraically in terms of
particle momentum and the center of mass coordinate $y$. This leads
to the self-consistent system for the variables $y,P$.

Let us find a closed system of equations for the dynamical variables
$y$, $P$. Condition (\ref{M-P}) is solved by the following ansatz
(initial form) for $M$:
\begin{equation}\label{MP}
    M=-\frac{s}{m}\gamma P\,,
\end{equation}
where $\gamma$ is an unknown dimensionless function. The function
$\gamma$ must be positive because the Frenkel angular momentum and
extended momentum are collinear for $s<0$ and anti-collinear for
$s>0$, see the comments to the formula (\ref{chir}). In the free
limit $A,F=0$, $\gamma=1$. The constraints (\ref{PMc-constr})
determine the current value of $\gamma$ in the presence of
interaction. On substituting the ansatz (\ref{MP}) into the
conditions (\ref{PMc-constr}), we obtain the cubic equation for
$\gamma$,
\begin{equation}\label{z-func}
    z\gamma^3+\gamma^2-1=0\,,\qquad z=gs(P,F)/m^3\,.
\end{equation}
The positive root that tends to $1$ in the limit $g\to0$ is
relevant. The trigonometric formula gives the following exact
solution for $\gamma$:
\begin{equation}\label{z-eq}
    \frac{1}{\gamma}=
    \cos\bigg(\frac{1}{3}\arcsin\frac{\sqrt{27}z}{2}\bigg)+\frac{1}{\sqrt{3}}\sin\bigg(\frac{1}{3}\arcsin\frac{\sqrt{27}z}{2}\bigg)\,.
\end{equation}
If the field $F$ is weak, the solution involves a small parameter
$z$. The Macloren expansions for the quantities $\gamma$ and
$1/\gamma$ read
\begin{equation}\label{z-decomp}
    \frac{1}{\gamma}=1+\frac{1}{2}z-\frac{3}{8}z^2+O(z^3)\,,\qquad
    \gamma=1-\frac{1}{z}z+\frac{5}{8}z^2+O(z^3)\,.
\end{equation}
Relation (\ref{MP}), (\ref{z-eq}) allow us to eliminate the variable
$M$ in terms of the mass center position $y$ and extended momentum
$P$. The final form of the equations of motion for $y$ and $P$ reads
\begin{equation}\label{EoM-cma}\begin{array}{c}\displaystyle
    \frac{1}{e}\frac{dy}{d\uptau}=\bigg(1-\frac{1}{2}\frac{(g-2)z\gamma}{g+(g+1)z\gamma}\bigg)P
    -\frac{s}{2m}\frac{g(g-2)\gamma(1+z\gamma)}{g+(g+1)z\gamma}F+
    \frac{s}{2m}\frac{g\gamma[P,\partial]z}{g+(g+1)z\gamma}\,,\\[7mm]\displaystyle
     \frac{1}{e}\frac{dP}{d\uptau}=-\frac{1}{2}m^2\gamma\partial
     z+\bigg[\frac{1}{e}\frac{dy}{d\uptau},\,F\bigg]\,.
\end{array}\end{equation}
Here, the functions $z$ and $\gamma$ are defined in (\ref{z-func}),
(\ref{z-eq}) (see (\ref{z-decomp}) for Macloren series expansion for
$\gamma$). The evolutionary equations (\ref{EoM-cma}) are
complimented by the mass shell constraint
\begin{equation}\label{constr-Pi}
    (P,P)+m^2(1+z\gamma(z))\approx0\,.
\end{equation}
By construction, relations (\ref{EoM-cma}), (\ref{constr-Pi})
determine a self-consistent system of equations describing the
classical trajectory of the mass center. The system
(\ref{EoM-cma-app}), (\ref{constr-Pi-app}) has no analog in the
higher space-time dimensions, where the particle spin has physical
degrees of freedom.

The system (\ref{EoM-cma-app}), (\ref{constr-Pi-app}) involves the
model parameters, mass and spin, in a complicated way. As the spin
of all real particles is proportional to the Plank constant $\hbar$
(we put $s=\sqrt{3/4}\hbar$ for electron), the quantity $z$ can be
considered as small parameter in practical applications, at least if
the electromagnetic field is not too strong. In this setting, it is
interesting to consider the spin as small parameter and derive
approximate equations of motion, which include contributions up
certain order in $s$ (or in $z$). Formula (\ref{z-decomp})
determines the decomposition of $\gamma$ up to the second order in
$z$. With the same precision, we get the following representations
for equations (\ref{EoM-cma}):
\begin{equation}\label{EoM-cma-app}\begin{array}{c}\displaystyle
    \frac{1}{e}\frac{dy}{d\uptau}=\Big(1-\frac{1}{2}\frac{g-2}{g}z+\frac{(g-2)(3g+2)}{4g^2}z^2\Big)P
    -\frac{(g-2)s}{2m}\Big(1-\frac{1}{2}\frac{g+2}{2g}z\Big)F-\\[7mm]\displaystyle
    -\frac{s}{2m}[P,\partial]z+O(s^3)
    \,,\\[7mm]\displaystyle
    \frac{1}{e}\frac{dP}{d\uptau}=-
    \frac{1}{2}m^2\Big(1-\frac{1}{2}z\Big)\partial z+\Big[\frac{1}{e}\frac{dy}{d\uptau},\,F\Big]+O(s^3)\,.
\end{array}\end{equation}
The constraint (\ref{constr-Pi}) takes the following form:
\begin{equation}\label{constr-Pi-app}
    (P,P)+m^2\Big(1+z-\frac{1}{2}z^2+O(s^3)\Big)\approx0\,.
\end{equation}
Equations (\ref{EoM-cma-app}), (\ref{constr-Pi-app}) account all the
possible corrections for the spinning particle classical trajectory
that are at least quadratic in $s$. The contributions, that involve
the electromagnetic field strength, account for the deviation of
spinning particle classical trajectory caused by the presence of
spin. The gradient contributions, which involve $\partial z$,
account the interaction between the particle magnetic moment and
external field. The case $s=0,z=0$ corresponds to the option of
scalar particle. If $z=0$, $s\neq0$ the particle has spin, but its
magnetic moment vanishes. Equations (\ref{EoM-cma-app}),
(\ref{constr-Pi-app}) tell us that the particle trajectories of
spinning and spinless particle are different even if the
gyromagnetic ratio is zero. This means that the spin has the
mechanism of influence on the particle path, which is not caused by
the coupling of interaction of magnetic moment and electromagnetic
field.

The classical trajectory of the mass center in the phase-space is a
curve. The classical trajectory of the particle with the initial
vales of the center of mass coordinate $y(0)$ and extended momenta
$P(0)$ is denoted as follows:
\begin{equation}\label{cma-path}
    y^\mu=\mathcal{Y}{}^\mu{}_{cl}(\uptau;y(0),P(0))\,,\qquad
    P=\mathcal{P}{}^\mu{}_{cl}(\uptau;y(0),P(0))\,,\qquad \mu=0,1,2.
\end{equation}
Here, the proper time $\uptau$ is the parameter on the curve; the
initial moment of time is $\uptau=0$. There four independent initial
data among the quantities $y(0)$, $P(0)$. At first, $y(0)$ and
$P(0)$ meet the mass shell constraint condition (\ref{constr-Pi}).
The constraint reduced the number of independent initial data by
one. The triviality of other initial data is connected to the gauge
invariance of equations (\ref{EoM-cma-app}). The particular choice
of the parameter on the curve has not no sense in the
reparametrization-invariant models. The fixation of initial point of
the particle trajectory corresponds to a particular gauge fixing,
which is a natural ambiguity. Hereinafter, we assume that
\begin{equation}\label{sp-Piy}
    (y(0),P(0))=0\,.
\end{equation}
In the absence of electromagnetic field, this condition implies that
the vector $y(0)$ connects the origin and mass center trajectory by
the shortest path (see relations (\ref{pJ-ny})). This interpretation
is not valid in the interacting theory because the vectors $\dot{y}$
and $P$ have different direction, in general. Relations
(\ref{M-Frenkel}), (\ref{M-P}), (\ref{z-func}), (\ref{z-eq}) express
the vectors $y(0)$, $P(0)$ in terms of initial vales of the momentum
and total angular momentum $p{}(0)$, $J{}(0)$ by the rule
\begin{equation}\label{pJ-Piy-int}
    p{}(0)=P{}(0)-A(y(0))\,,\qquad
    J{}(0)=\bigg[y{}(0),P{}(0)-A(y(0))\bigg]-\frac{s}{m}\gamma(z(0))P(0)\,.
\end{equation}
The notation $z(0)$ is used for initial value of the quantity $z$
(\ref{z-eq}). The solution of equations (\ref{pJ-Piy-int}) with
respect to $y(0)$ and $P(0)$ expresses the initial data in terms of
the values of momentum and total angular momentum. It is not
possible to express $y(0)$ and $P(0)$ explicitly. An approximate
representation is obtained by the fixed-point iteration method for
the system of equations
\begin{equation}\label{}
    P{}(0)=p(0)+A(y(0))\,,\qquad  y(0)=\frac{1}{1+z(0)\gamma(z(0))}\bigg[\frac{P(0)}{m},
    \frac{J(0)}{m}\bigg]\,.
\end{equation}
The free solution (\ref{pJ-ny}) determines the zero approximation
for unknowns $y(0)$, $P(0)$. As we see, the position of the
classical trajectory of spinning particle mass center contains all
the valuable information about the dynamics of the model.

Let us now consider the particle motion in the position
representation. The constraints (\ref{constr-pos-rep-int}) imply
that the difference vector $\omega$ (\ref{d-vec}) is orthogonal the
extended momentum and normalized in each moment of proper time (see
the relations $\Theta_2,\Theta_4\approx0$). Condition (\ref{d-vec})
expresses the difference vector in terms of the particle coordinate
center of mass position. Relation (\ref{cma-path}) determines the
classical trajectory of mass center in the parametrical form.
Combining (\ref{d-vec}), (\ref{cma-path}) with the constraints
$\Theta_2,\Theta_4\approx0$ (\ref{constr-pos-rep-int}), we get the
following relations for particle position $x$ and proper time
$\uptau$:
\begin{equation}\label{x-yP}
    (x-\mathcal{Y}{}^\mu{}_{cl},\mathcal{P}{}^\mu{}_{cl})=0\,,\qquad
    (x-\mathcal{Y}{}^\mu{}_{cl},x-\mathcal{Y}{}^\mu{}_{cl})-r^2=0\,.
\end{equation}
These equations determine the world sheet in implicit form. For each
particular value of $\uptau$, relations (\ref{x-yP}) tell us that
the particle position $x$ lies on a circle of radius $r$, with the
extended momentum $P$ being the normal to the plane. The vector
$\mathcal{Y}_{cl}$ determines the position of the circle center.
With the proper time being the classical position of the mass center
moves along its trajectory (\ref{cma-path}). The set of possible
particle positions moves along a cylindrical hypersurface that
surrounds the mass center classical trajectory. This hypersurface
coincides with a spinning particle world sheet. In the explicit
form, the world sheet equation for the particle coordinates follows
from the system (\ref{x-yP}) if the proper time variable $\uptau$ is
expressed in terms of current particle position $x$. In the case of
free particle,
\begin{equation}\label{}
\phantom{\frac12}\mathcal{Y}{}^\mu{}_{cl}(\uptau;y(0),p(0))=y^\mu(0)+p^\mu(0)\uptau\,,\qquad
\mathcal{P}{}^\mu{}_{cl}(\uptau;y(0),p(0))=p^\mu(0)\,,\phantom{\frac12}
\end{equation}
and the solution reads
\begin{equation}\label{t-xP}
    \phantom{\frac12}\uptau=-\frac{(x,p(0))}{m^2}\,.\phantom{\frac12}
\end{equation}
The world sheet of spinning particle, being determined by the
relations (\ref{x-yP}), is the hypercylinder (\ref{ws}). In all the
instances, the hypersuface (\ref{x-yP}) represents a set of all
possible spinning particle positions with the fixed values of
momentum and total angular momentum. The classical trajectories of
the spinning particle are general lines on the world sheet. This
result demonstrates that concept of the world sheet of spinning
particle is viable at the interacting level. Another important
observation is that the spinning particle trajectory can deviate
from the smooth curve, being the center of mass path. This means
that the zitterbewegung phenomenon is observed at the interacting
level.

\section{Motion in the uniform electric field}

In this section, we describe the particle motion in the uniform
electric field. The vector potential and field strength are chosen
in the following form:
\begin{equation}\label{E-vec}
    A(x)=(Ex^1,0,0)\,,\qquad F=(0,0,E)\,,
\end{equation}
where $x^1$ is the space-time coordinate and $E$ is the field
strength. The field is termed ''electric'' because only time
component of the vector potential is nonzero. In the first part of
this section, we find the mass center classical trajectory. The
motion in the direction of field strength vector is unform. Its
projection of the trajectory onto the orthogonal to the field
strength vector plane is shown to be hyperbola. The shape of the
classical path depends on spin, so the trajectories of spinning and
scalar particles are different. In the second part of the section,
we derive the spinning particle world sheet. If the vector of
initial velocity is orthogonal to the fields strength, the world
sheet is pseudotorus with the inner radius $r$. The outer radius of
pseudotorus is determined by the particle mass, particle spin,
electric field strength.

Let us first determine the center of mass classical trajectory. In
the case of unform electric field (\ref{E-vec}), the equations of
motion (\ref{EoM-cma}) read
\begin{equation}\label{EoM-Hom}\begin{array}{c}\displaystyle
    \frac{1}{e}\frac{dy^0}{d\uptau}=\bigg(1-\frac{1}{2}\frac{(g-2)z\gamma(z)}{g+(g+1)z\gamma(z)}\bigg)P^0\,,\qquad
    \frac{1}{e}\frac{dy^1}{d\uptau}=\bigg(1-\frac{1}{2}\frac{(g-2)z\gamma(z)}{g+(g+1)z\gamma(z)}\bigg)P^1\,,
    \\[7mm]\displaystyle
    \frac{1}{e}\frac{dy^2}{d\uptau}=\bigg(1-\frac{1}{2}\frac{(g-2)z\gamma(z)}{g+(g+1)
    z\gamma(z)}\bigg)P^2-
    \frac{1}{2}\frac{s}{m}\frac{g(g-2)(1+z\gamma(z))\gamma(z)}{g+(g+1)z\gamma(z)}E\,,
\end{array}\end{equation}
\begin{equation}\label{EoM-Hom-2}\begin{array}{c}\displaystyle
    \frac{1}{e}\frac{dP^0}{d\uptau}=-\Big(1-\frac{1}{2}\frac{(g-2)z\gamma(z)}{g+(g+1)z\gamma(z)}\Big)
    EP^1\,,\phantom{a}
    \frac{1}{e}\frac{dP^1}{d\uptau}=-\Big(1-\frac{1}{2}\frac{(g-2)z\gamma(z)}{g+(g+1)z\gamma(z)}\Big)EP^0\,,
    \\[7mm]\displaystyle
    \frac{1}{e}\frac{dP{}^2}{d\uptau}=0\,.
\end{array}\end{equation}
Here, the variable $z$ denotes a dimensionless combination
(\ref{z-func}) of the extended momentum $P$, field strength $E$,
gyromagnetic ratio $g$, mass $s$ and spin $s$. The univariate
function $\gamma(z)$ is defined in (\ref{z-eq}). Equation
(\ref{z-decomp}) determines two leading orders of decomposition of
$\gamma$ in the small parameter $z$. The dynamical variables are the
center of mass position coordinates $y{}^\mu$ and components
extended momenta $P{}^\mu$. The evolutionary equations are
supplemented by the constraint  (\ref{constr-Pi}). The arbitrary
function $e=e(\uptau)$ denotes einbein, and $\uptau$ is the proper
time. We are interested in the solution $y(\uptau)$, $P(\uptau)$ to
the equations (\ref{EoM-Hom}), (\ref{EoM-Hom-2}) with the initial
condition $y(\uptau)|_{\uptau}=y(0)$,
$P(\uptau)|_{\uptau=0}=P(0)$.

Equations (\ref{EoM-Hom}), (\ref{EoM-Hom-2}) have two obvious
integrals of motion $(P,P)=\text{const}$ and $(P,E)=\text{const}$.
This implies conservation of quantities $z$ and $\gamma$
(\ref{z-func}), (\ref{z-eq}). With account of this fact, equations
(\ref{EoM-Hom}), (\ref{EoM-Hom-2}) reduce to a linear system with
respect to the dynamical variables $y^\mu$, $P^\mu$. The motion is
uniform in the direction of field. The projection of the center of
mass velocity is determined by the value of momentum component of
momentum $P^2(0)$, being connected to the auxiliary quantity $z(0)$
(\ref{z-func}). If the spinning particle center of mass classical
trajectory lies in the plane orthogonal to the field, the equation
$\dot{y}{}^2=0$ determines the selected value of momentum projection
$\underline{P}^2(0)$ of constant $\underline{z}(0)$,
\begin{equation}\label{z-bar}
\underline{P}{}^2(0)=\frac{(g-2)s}{2m}E+O(s^2)\,,\qquad
\underline{z}(0)=\frac{(g-2)s^2}{2gm^6}E^2+O(s^3)\,.
\end{equation}
It is interesting to mention that the first-order correction for
$\underline{P}^2(0)$ vanishes for the particle without anomalous
magnetic moment. In the gauge
\begin{equation}\label{}
    \frac{1}{e}=\bigg(1-\frac{1}{2}\frac{(g-2)z\gamma(z)}{g+(g+1)z\gamma(z)}\bigg)\,,
\end{equation}
the solution to the system (\ref{EoM-Hom}), (\ref{EoM-Hom-2}) with
the initial condition reads
\begin{equation}\label{y0y1-path}
    y^0=y^0(0)\cosh\frac{\uptau}{E}+y^1(0)\sinh\frac{\uptau}{E}+\frac{P^1(0)}{E}\,,\quad
    y^1=y^0(0)\sinh\frac{\uptau}{E}+y^1(0)\cosh\frac{\uptau}{E}+\frac{P^0(0)}{E}\,,
\end{equation}
\begin{equation}\label{y2-path}
    y^2=y^2(0)+\bigg(P^2(0)-\frac{m^2z(0)}{P^2(0)}\frac{(g-2)(1+z(0)
    \gamma(z(0)))\gamma(z(0))}{2g+(g+4)z(0)\gamma(z(0))}\bigg)\uptau.
\end{equation}
For the extended momentum $P^{\mu}$, the classical trajectory reads
\begin{equation}\label{P0-path}
    P^0=P^0(0)-Ey^0(0)\sinh\frac{\uptau}{E}+Ey^1(0)(\cosh\frac{\uptau}{E}-1)\,,
\end{equation}
\begin{equation}\label{P1P2-path}
    P^1=P^1(0)-Ey^0(0)(\cosh\frac{\uptau}{E}-1)+Ey^1(0)\sinh\frac{\uptau}{E}\,,\qquad
    P^2-P^2(0)=0\,.
\end{equation}
We suppose that the initial position $y(0)$ and momentum satisfy
constraint (\ref{constr-Pi}) and condition (\ref{sp-Piy}). Formula
(\ref{pJ-Piy-int}) expresses the initial data $y(0)$, $P(0)$ in
terms of the initial values of momentum and total momentum $p(0)$,
$J(0)$. The exact solution for the variables $y^\mu(0),\mu=0,1,2$
reads
\begin{equation}\label{y-uni}\begin{array}{c}\displaystyle
    y{}^0(0)=-\frac{1}{m^2}\frac{\epsilon^{ij}p{}^i(0)J{}^i(0)}{1+z(0)\gamma(z(0))}\,,\\[7mm]\displaystyle
    y{}^1(0)=\frac{1}{m^2}\frac{p{}^0(0)J{}^2(0)-p{}^2(0)J{}^0(0)}{1+z(0)\gamma(z(0))}
    \Big(1-\frac{1}{m^2}\frac{EJ{}^2(0)}{1+z(0)\gamma(z(0))}\Big)^{-1}\,,\\[7mm]\displaystyle
    y{}^2(0)=\frac{1}{m^2}\frac{p{}^1(0)J{}^0(0)-p{}^0(0)J^1(0)}{1+z(0)\gamma(z(0))}-\frac{1}{m^4}
\frac{EJ{}^1(0)}{1+z(0)\gamma(z(0))}\frac{p{}^0(0)J{}^2(0)-p{}^2(0)J{}^0(0)}{1+z(0)\gamma(z(0))}\times\\[7mm]\displaystyle
\times\Big(1-\frac{1}{m^2}\frac{EJ{}^2(0)}{1+z(0)\gamma(z(0))}\Big)^{-1}\,.
\end{array}\end{equation}
The initial value of the extended momentum $P{}^\mu(0),\mu=0,1,2$ is
expressed as follows:
\begin{equation}\label{P-uni}\begin{array}{c}\displaystyle
    P{}^0(0)=p{}^0(0)+\frac{E}{m^2}\frac{p{}^0(0)J{}^2(0)-p{}^2(0)J{}^0(0)}{1+z(0)\gamma(z(0))}
    \Big(1-\frac{1}{m^2}\frac{EJ{}^2(0)}{1+z(0)\gamma(z(0))}\Big)^{-1}\,,\\[7mm]\displaystyle
    P{}^i(0)=p{}^i(0)\,,\qquad i=1,2\,.
\end{array}\end{equation}
Relations (\ref{y0y1-path})-(\ref{P-uni}) determine the classical
trajectory of the mass center of the particle with the prescribed
initial values of momentum $p(0)$ and total angular momentum $J(0)$.

Let us explain the geometric meaning of equations
(\ref{y0y1-path})-(\ref{P1P2-path}). Relations (\ref{y0y1-path}),
(\ref{y2-path}) determine the classical trajectory of mass center of
the particle in space-time. The classical trajectory
(\ref{P0-path}), (\ref{P1P2-path}) of extended momentum is
consequence of definition of $P$ in terms of particle velocity. Due
to this fact, the space-time trajectory of mass center contains all
the valuable information about the dynamics of the model. The
initial values of momentum and total angular momentum determine the
positions of mass center classical path (\ref{y0y1-path}),
(\ref{y2-path}) in space-time by the rule (\ref{y-uni}),
(\ref{P-uni}). Equation (\ref{y0y1-path}) tells us that the
projection of the mass center path onto the orthogonal to the fields
strength vector plane is a hyperbola. The center of symmetry has the
coordinates
\begin{equation}\label{yP}
    \widehat{y}^0(0)=y^0(0)+\frac{P^1(0)}{E}\,,\qquad
    \widehat{y}^1(0)=y^1(0)+\frac{P^0(0)}{E}\,.
\end{equation}
The focal distance $\Phi$ of hyperbola is the function of momentum
projection onto the direction of field,
\begin{equation}\label{Phi}
\Phi(z(0))=\frac{m}{E}\Big(1+z(0)\gamma(z(0))+
\frac{m^4E^2}{g^2s^2}z^2(0)\Big)^{1/2}\,.
\end{equation}
The focal distance depends on the value of spin, so the shape of
paths of scalar and spinning particle is different. Equation
(\ref{y2-path}) determines the deviation of the mass center
trajectory in the direction of field. The path (\ref{y0y1-path}),
(\ref{y2-path}) is planar curve if the condition $\dot{y}{}^2=0$. In
this case, the quantities $\widehat{y}^0$, $\widehat{y}^1$
(\ref{yP}) and $y^2(0)$ serve as the coordinates of hyperbola
center. The focal distance of hyperbola reads
\begin{equation}
\Phi(\underline{z})=\frac{m}{E}\Big(1+\frac{s^2}{m^4}\frac{(g-2)(3g+2)}{8}E^2+O(s^3)\Big)\,.
\end{equation}
Here, we note that the corrections to the focal distance are at
least quadratic in spin. In the first order in spin, the classical
paths of spinning and scalar particle, with the initial velocity
vector being orthogonal to the field strength, have one and the same
shape. In the second-order approximation in $s$, the classical paths
of spinning and scalar particles are different.

The world sheet of spinning particle in the uniform electric field
is determined by the conditions (\ref{x-yP}), where equations
(\ref{y0y1-path}), (\ref{y2-path}) describe the classical trajectory
of mass center. As in the general case, we obtain the system of two
equations for the unknown vector $x$ and parameter on the curve
$\uptau$. A single equation for the particle coordinate follows from
this system if the variable $\uptau$ is expressed from system in
terms of the positions $x$, and initial data $y(0)$, $P(0)$ (or
$p(0)$, $J(0)$). It is not possible to express $\uptau$ from the
system (\ref{x-yP}), (\ref{y0y1-path}), (\ref{y2-path}) for general
initial data explicitly, so the world sheet is determined in the
implicit form. The general result of the previous paragraph tells us
that it a cylindrical surface of radius $r$ surrounding the mass
center classical trajectory. The problem of world sheet construction
simplifies in the case $\dot{y}=0$ of motion in the orthogonal to
the field direction plane. The system (\ref{y0y1-path}),
(\ref{y2-path}) becomes algebraic with respect to the unknowns $x$,
$\uptau$. The quantity $\uptau$ is expressed from equations by the
application of the techniques of resultants. The world sheet
equation eventually reads
\begin{equation}\label{e-ws}\begin{array}{c}\displaystyle
(-(x{}^0-\widehat{y}{}^0)^2+(x{}^1-\widehat{y}{}^1)^2+(x{}^2-\widehat{y}{}^2)^2+\Phi^2(\underline{z})-r^2)^2+\\[5mm]\displaystyle
+4\Phi^2(\underline{z})(-(x{}^0-\widehat{y}{}^0)^2+(x{}^1-\widehat{y}{}^1)^2\,)=0\,,
\end{array}\end{equation}
where $\Phi(\underline{z})$ is determined in (\ref{z-bar}),
(\ref{Phi}). The hypersurface, being defined by equation
(\ref{e-ws}), is a pseudotorus of external radius
$\Phi(\underline{z})$ and internal radius $r$. The constant vector
$\widehat{y}(0)=( \widehat{y}^0(0), \widehat{y}^1(0), y^2(0))$
(\ref{yP}) denotes the position of symmetry center of pseudotorus.
The vector of electric field strength (\ref{E-vec}) determines the
direction of symmetry axis of pseudotorus. Equations (\ref{z-bar}),
(\ref{yP}), (\ref{y-uni}), (\ref{P-uni}) express the components
$y^\mu(0),\mu=0,1,2,$ in terms of initial values of the momentum and
total angular momentum $p(0)$, $J(0)$. This final result
demonstrates that spinning particles trajectories can be considered
as curves on the set of hypersurfaces even in the presence of
external electromagnetic field.

\section{Conclusion}

In this article, we have studied an issue of description of dynamics
of spinning particle by means of recently proposed world sheet
concept. We have considered the class of massive spinning particle
models in three-dimensional Minkowski space whose quantization
corresponds to the irreducible representation of the Poincare group.
The classical path of the free particle lie on a hypercylinder with
the time like axis. The vector of momentum determines the direction
of cylinder axis. The position of the hypercylinder in space-time is
determined by the value of total angular momentum. The radius of
cylinder is determined by the representation. All the paths that lie
on one and the same cylinder are connected by the gauge
transformations.

Using the ideas of the papers \cite{Univ,Deriglazov14}, we have
proposed a constructive procedure of inclusions of consistent
interactions between the particle and general electromagnetic field.
The coupling is not minimal. The electromagnetic field potential and
its strength are taken not at the particle position, but in a
special point with reduced gauge symmetry, termed a particle mass
center. As we see, such reinterpretation of the dynamical variables
preserves gauge invariance and ensures that the gauge equivalence of
the particle paths forms a cylindrical hypersurface in Minkowski
space. The axis of cylinder is associated with the path of point
massive particle, whose dynamics is described by the equations of
the papers \cite{Univ,Deriglazov14}.

The specifics of three-dimensional model is that the spin has no
physical degrees of freedom. In this case, the center of mass
classical trajectory contains all the valuable information about the
dynamics of the particle.  We derive a system of ordinary
differential equations that describe the center of mass path. The
dynamical variables of this system are classical position of the
mass center and conjugated momentum, while all internal space
variables are excluded. This system has no analogs in higher
space-time dimensions. Assuming that the mass center trajectory is
known, we propose the procedure of construction of a hypersurface
that includes all the particle paths.

The general constructions are illustrated in the case of uniform
electric field. The particle's mass center equations of motion are
explicitly solved for general mass and spin.  If the center of mass
velocity vector is orthogonal to the field strength vector, the
center of mass path is a hyperbola. The focal distance of hyperbola
depends on mass, spin, and electric field strength. The world sheet
of the spinning particle is a pseudotorus whose outer radius is
determined by the particle mass and electric field strength. The
inner radius of pseudotorus equals to the cylinder, which is the
free particle world sheet. The relationship between the pseudotorus
parameters and initial values of the particle momentum and total
angular momentum is explicitly established.

\section{Acknowledgments}
The authors thank S.L. Lyakhovich, A.A. Sharapov and A.A. Deriglazov
for valuable discussions of this work and comments on the
manuscript. The was supported by RFBR, project number 20-32-70023.
The work of IAR was partially supported by the Foundation for the
Advancement of Theoretical Physics and Mathematics "BASIS".

\noindent E-mail address: dsc@phys.tsu.ru, retuntsev.i@phys.tsu.ru.

\end{document}